\documentstyle[preprint,aps]{revtex} 
\begin{document}
\input{epsf}
 
\preprint{UMHEP-433}
\draft
\title{Final state rescattering as a contribution to  
$B \to \rho \gamma$. }
\author{John F. Donoghue, Eugene Golowich and
Alexey A. Petrov}
\address{Department of Physics and Astronomy,
University of Massachusetts \\
Amherst MA 01003 USA}
\maketitle
\begin{abstract}
\noindent
We provide a new estimate of the long-distance component to the radiative 
transition $B \to \rho \gamma$.  Our mechanism involves the soft-scattering of
on-shell hadronic products of nonleptonic $B$ decay, as in the chain 
$B \to \rho\rho \to \rho\gamma$. 
We employ a phenomenological fit to scattering data 
to estimate the effect.  The specific intermediate states considered 
here modify the $B \to \rho \gamma$ decay rate at roughly the $5 \to 8\%$
level, although the underlying effect has the potential to be larger.  
Contrary to other mechanisms of long distance physics which have been
discussed in the literature, this yields a non-negligible modification of
the $B^0 \to \rho^0 \gamma$ channel and hence will provide an 
uncertainty in the extraction of $V_{td}$.  
This mechanism also affects the isospin relation between the rates for 
$B^- \to \rho^-\gamma$ and $B^0 \to \rho^0 \gamma$ and may 
generate CP asymmetries at experimentally observable levels.

\end{abstract}

\section{Introduction}

The study of radiative rare decays
of $B$ mesons can provide valuable information on certain 
parameters in the CKM matrix.  In particular, 
$B \to \rho \gamma$ is thought to be an especially clean 
mode for extracting the $V_{td}$ 
matrix element.$^{\cite{vtd}}$  Although not yet observed, 
this mode should be accessible for study at B-factories.  However, 
the smallness of the short-distance $B \to \rho \gamma$ 
amplitude raises the concern that the experimental signal could be 
influenced by long-distance effects, with the possibility of 
their contributing a non-negligible fraction to the decay rate.  

The first studies$^{\cite{gp}}$ of long distance contributions 
to rare $B$ decays employed the vector meson dominance (VMD) model 
({\it cf} Fig.~1(a)).  As applied to the $B \to \rho \gamma$ 
transition,$^{\cite{cheng}}$ this involves the decay $B \to \rho 
V^{*0}$ ($V^{*0}$ represents an off-shell neutral vector meson 
such as $\rho^0, \omega, \phi, J/\psi, \dots$) 
with subsequent conversion of the vector meson 
to the photon. More recently, light-cone QCD sum-rules have been 
used to analyze the weak annihilation long-distance contribution to 
$B \to \rho \gamma$.$^{\cite{lcsr}}$   Here, the main effect 
arises from direct emission of the final state photon from the 
light spectator quark in the $B$ meson followed by ${\bar b}q$ 
weak annihilation.   

Both these approaches suffer from some degree of theoretical 
uncertainty. Besides the usual model dependence of these methods,
in both cases no allowance is made 
for final state interactions (FSI). 
However, the $\rho\gamma $ decay mode can be generated by the decay of
a B meson into a hadronic final state which then rescatters into
$\rho\gamma $. It has recently been 
shown$^{\cite{we}}$ that soft-FSI effects are 
${\cal O}(1)$ in the large $m_b$ limit and cannot legitimately 
be ignored.  

In this paper, we provide an estimate of a 
long distance component to $B \to \rho \gamma$ which 
deals solely with on-shell transition amplitudes while explicitly taking 
the occurrence of FSI into account.  Our mechanism 
can be viewed as a unitarity analysis, where a $B$ decays 
into some intermediate state which then 
undergoes soft-rescattering into the final $\rho \gamma$ 
configuration.  Of course, we are not able to account for all possible
intermediate states, so that in order to provide a rough estimate for the
the effect we must analyse a few specific contributions.
For definiteness, we consider the 
$\rho^0 \rho^0$ and $\rho^+ \rho^-$ intermediate states 
in $B^0$ decay.  In the language of Regge
theory,$^{\cite{we,collins}}$  
the first of these proceeds by Pomeron exchange and is 
technically the dominant contribution, remaining nonzero 
in the heavy quark limit.  However, the 
$\rho^+ \rho^-$ contribution, whose rescattering is 
mediated by the $\rho$ trajectory and is thus nonleading, 
can be numerically important because $B^0 \to \rho^+\rho^-$ 
is color-allowed (whereas $B^0 \to \rho^0\rho^0$ is 
color-suppressed) and because the relatively low value of the $B$ mass 
turns out to blur somewhat the distinction between Regge-leading 
and nonleading contributions.  Finally, 
since the corrections considered here are not themselves 
proportional to $V_{td}$, their presence constitutes a potential 
source of serious error in the phenomenological extraction 
of $V_{td}$.$^{\cite{joao}}$  This underscores the importance 
of obtaining quantitative estimates of such effects.  

\section{Soft hadronic rescattering}
\noindent
Effects of final state interactions ({\it e.g.} Fig.~1(b)) are 
naturally described using the unitarity property of the 
${\cal S}$-matrix, ${\cal S}^\dagger {\cal S} = 1$.  
This condition implies that the ${\cal T}$-matrix, 
${\cal S} = 1 + i {\cal T}$, obeys
\begin{equation}
{\cal D}isc~{\cal T}_{B \rightarrow f} \equiv {1 \over 2i}
\left[ \langle f | {\cal T} | B \rangle -
\langle f | {\cal T}^\dagger | B \rangle \right]
= {1 \over 2} \sum_{I} \langle f | {\cal T}^\dagger | I \rangle
\langle I | {\cal T} | B \rangle  \ .
\label{unit}
\end{equation}
For the $B \to \rho \gamma$ transition the contribution 
from the $\rho$ intermediate state is the one which is most 
amenable to direct analysis, and we shall detail the 
$B \to \rho\rho \to \rho\gamma$ component throughout this 
paper.\footnote{One should keep in mind, however, that 
for the $S$-matrix to be unitary, inelastic effects 
such as diffractive dissociation (decay of the $B$ into a 
$\rho$ and a jet of particles which recombine into the $\rho \gamma$
final state) or any other intermediate state with suitable quantum 
numbers must also be present.$^{\cite{we}}$}   This encompasses 
the $\rho^0 \rho^0$ and $\rho^+\rho^-$ 
intermediate states for $B^0$ decay and $\rho^+ \rho^0$ for $B^+$ decay.  

The final state interaction of Fig.~1(b) together with 
the unitarity condition implies a discontinuity 
relation for the invariant amplitude ${\cal M}_{B \to \rho \gamma}$, 
\begin{eqnarray}
\lefteqn{2Disc~{\cal M}_{B \to \rho \gamma}^{\eta \theta}  
\epsilon^*_\eta (\lambda_3) \epsilon^*_\theta (\lambda_4)=}\nonumber
\\
& & \int \frac{d^4 p}{(2 \pi)^4} (2 \pi)^2 ~  d_{\mu \nu}^{(1)} ~
\delta (p^2 - m^2) 
d_{\alpha \beta}^{(2)} ~ \delta ((p_B-p)^2 - m^2) 
{\cal M}_{B \to \rho \rho}^{\mu \alpha}
{\cal M}_{\rho \rho \to \rho \gamma}^{* ~ \nu \beta, \eta \theta} ~  
\epsilon^*_\eta (\lambda_3) \epsilon^*_\theta (\lambda_4) \nonumber \\
& & = \sum_{\lambda_1, \lambda_2}
\int \frac{d^4 p}{(2 \pi)^4} (2 \pi)^2 \delta (p^2 - m^2)
\delta ((p_B-p)^2 - m^2) \nonumber \\ 
& & \phantom{xxxxxx}\times {\cal M}_{B \to \rho \rho}^{\mu \alpha} ~
\epsilon^*_\mu (\lambda_1) \epsilon^*_\alpha (\lambda_2) 
~ \cdot 
\epsilon^*_\nu (\lambda_1) \epsilon^*_\beta (\lambda_2) ~
{\cal M}_{\rho \rho \to \rho \gamma}^{* ~ \nu \beta, \eta \theta} ~
\epsilon^*_\eta (\lambda_3) \epsilon^*_\theta (\lambda_4) \ ,
\end{eqnarray}
with $d^{(i)}_{\mu \nu}$ being the polarization tensors of
$\rho$ mesons in the loop.  When expressed 
in terms of helicity amplitudes, the above discontinuity 
formula simplifies to 
\begin{eqnarray}
\lefteqn{2Disc~ {\cal M}_{B \to \rho \gamma} (\lambda_3 \lambda_4)=} 
\nonumber \\
& & \sum_{\lambda_1 \lambda_2}
\int \frac{d^4 p}{(2 \pi)^4} (2 \pi)^2 \delta (p^2 - m_\rho^2)
\delta ((p_B-p)^2 - m_\rho^2) 
{\cal M}_{B \to \rho \rho}(\lambda_1 \lambda_2) 
{\cal M}^*_{\rho \rho \to \rho \gamma} (\lambda_1 \lambda_2;
\lambda_3 \lambda_4)  \ \ .
\end{eqnarray}
The FSI itself will arise from the scattering $\rho\rho \to \rho
\gamma$.  We shall employ the Regge-pole description for this 
rescattering process, requiring both Pomeron and 
$\rho$-trajectory contributions.  

To begin, however, we recall from Regge phenomenology the 
well-known invariant amplitude for the scattering of particles 
with helicities $\{ \lambda_i\}$,$^{\cite{collins}}$ 
\begin{eqnarray}
{\cal M}_{i \rightarrow f}^{\lambda_1 \lambda_2;
\lambda_3 \lambda_4}=
-\left( \frac{-t}{s_0} \right)^{m/2}~
\frac{e^{-i \pi \alpha(t)} + {\cal J}} {2 \sin \pi \alpha(t)}
\gamma_{\lambda_3 \lambda_4}^{\lambda_1 \lambda_2}
\left( {s \over s_0} \right)^{\alpha (t)}\ \ ,
\label{regge}
\end{eqnarray}
where $m=|\lambda_3 - \lambda_1| + |\lambda_4 - \lambda_2|$ and 
$|{\cal J}| = 1$. In the $B \to \rho\rho$ weak decay, 
the $\rho \rho$ state can exist in any of three helicity 
configurations, $\lambda_1 = \lambda_2 = +1$, $\lambda_1 = 
\lambda_2 = -1$ and $\lambda_1 = \lambda_2 = 0$ (or in 
obvious notation $++$, $--$ and $00$).  Moreover, the Pomeron 
(and near $t = 0$ also the leading $\rho$) exchange does not 
change the helicities 
of the rescattering particles, {\it i.e.} $\lambda_1=\lambda_3,
\lambda_2=\lambda_4$.  In view of this and noting that the 
photon helicity must have $|\lambda_\gamma|= 1$, we omit 
the $00$ helicity configuration from further consideration.  
This is equivalent to maintaining the condition of gauge invariance.  
Moreover, since parity invariance constrains the $++$ and $--$ 
helicity contributions to be equal, we drop the 
$\lambda_1\lambda_2$ superscript hereafter and take 
for the residue couplings 
$\gamma_{--}^{--} = \gamma_{++}^{++} \equiv \gamma$. 
Throughout we use the linear trajectory forms, 
\begin{equation}
\alpha_P (t) = \alpha_P^0 + \alpha_P' t \ , \qquad 
\alpha_\rho (t) = \alpha_\rho^0 + \alpha_\rho' t \ \ .
\label{trajs}
\end{equation}

In the following discussion, we limit 
ourselves for clarity's sake to $B^0 \to \rho^0 \gamma$ decay.  
The Pomeron part of the FSI occurs for the $\rho^0\rho^0$ 
intermediate state. We obtain 
\begin{eqnarray}
Disc~ {\cal M}^{(P)}_{ B^0 \to \rho^0 \gamma} &=& \frac{1}{16 \pi s} 
{\cal M}_{B^0 \to \rho^0 \rho^0} \int^0_{t_{min}} \ dt \ 
{\cal M}_{\rho^0 \rho^0 \to \rho^0 \gamma}^* \nonumber \\
&=& -{\gamma_P \over 32 \pi s} \left( {s\over s_0}
\right)^{\alpha_P^0}{\cal M}_{B^0 \to \rho^0 \rho^0} 
\int^0_{t_{min}} \ dt \ e^{\alpha_P' \ln(s/s_0)t} ~
\left( \cot[\pi\alpha (t)/2] + i \right) \ \ ,
\label{pomint}
\end{eqnarray}
where ${\cal J}=+1$ for Pomeron exchange.  Observe that 
the cotangent factor will diverge at some $t = t_0$ such that 
$\alpha (t_0) = 0$.  For the Pomeron, this occurs at 
$t_0 \simeq -4.3$~GeV$^{-2}$, which lies outside the 
forward diffraction peak.  Such spurious behavior 
has been well-known since the early days of Regge phenomenology 
and is avoidable by restricting the range of integration to the 
diffraction peak, by employing a modified `phenomenological' amplitude
or by invoking daughter trajectories to cancel the divergence.
We adopt the first of these procedures and find 
\begin{eqnarray}
Disc~ {\cal M}^{(P)}_{ B^0 \to \rho^0 \gamma} 
&=& - {\gamma_P \over 32 \pi} {s_0 \over \alpha_P'\ln(s/s_0)}  
(0.26 - 0.92~i) \left( {s \over s_0}\right)^{\alpha_P^0 - 1}
{\cal M}_{ B^0 \to \rho^0 \rho^0 } \nonumber \\
&\equiv& \epsilon_P \left(\frac{s}{s_0}\right)^{\alpha_P^0-1}
{\cal M}_{ B^0 \to \rho^0 \rho^0 } \ \ , 
\label{disk}
\end{eqnarray}
where $s_0 \simeq 1$~GeV. The numerical quantity $\epsilon_P$ 
is defined to encode the strength of the Pomeron-mediated rescattering. 
We discuss later how to determine the Pomeron residue function 
$\gamma_P$ by fitting to experimental data.  

The quantity $Disc~ {\cal M}^{(P)}_{ B^0 \to \rho^0 \gamma}$ 
must itself be inserted as input to a dispersion relation 
for ${\cal M}^{(P)}_{ B^0 \to \rho^0 \gamma}$.  Since 
the $s$-dependent factor in Eq.~(\ref{disk}) is almost 
constant, the dispersion relation for the Pomeron contribution 
will require a single subtraction. Approximating 
$\alpha_P^0 \simeq 1$, we have 
\begin{eqnarray} \label{pom}
{\cal M}^{(P)}_{ B^0 \to \rho^0 \gamma}(m_B^2) &=&
{\cal M}^{(P)}_{ B^0 \to \rho^0 \gamma}(0) +
\frac{\epsilon_P}{\pi}
{\cal M}_{ B^0 \to \rho^0 \rho^0 } m_B^2
\int_{4 m_\rho^2}^\infty \ \frac{ds}{s(s-m_B^2)} \nonumber \\
&=& {\cal M}^{(P)}_{ B^0 \to \rho^0 \gamma}(0) +
\frac{\epsilon_P}{\pi}
{\cal M}_{ B^0 \to \rho^0 \rho^0 } 
\ln \Bigl(1 - \frac{m_B^2}{4 m_\rho^2}\Bigr) \ \ .
\end{eqnarray}

We now turn to the process $B^0 \to \rho^+ \rho^- \to \rho^0 \gamma$.  
The formulas derived for the leading Pomeron contribution of 
Eq.~(\ref{disk}) are readily applicable to this case if one 
replaces the Pomeron trajectory by the $\rho$ trajectory, 
but now with ${\cal J}= -1$.  This yields for the discontinuity function 
\begin{eqnarray} \label{diskrho}
Disc~ {\cal M}^{(\rho)}_{ B^0 \to \rho^0 \gamma} 
&=& -{\gamma_\rho \over 16 \pi s} \left( {s\over s_0}
\right)^{\alpha_\rho^0}{\cal M}_{B^0 \to \rho^+ \rho^-} 
\int^0_{t_{min}} \ dt \ e^{\alpha_\rho' \ln(s/s_0)t} 
\left( 1 + i \tan [\pi\alpha (t)/2] \right) \ \ , \nonumber \\
&=& - {\gamma_/rho \over 32 \pi} {s_0 \over \alpha_\rho'\ln(s/s_0)}  
(0.92 - 0.33~i) \left( {s \over s_0}\right)^{\alpha_\rho^0 - 1}
{\cal M}_{ B^0 \to \rho^0 \rho^0 } \nonumber \\
&\equiv& \epsilon_\rho \left(\frac{s}{s_0}\right)^{\alpha_\rho^0-1}
{\cal M}_{ B^0 \to \rho^+ \rho^- } \ \ , 
\end{eqnarray}
where $\epsilon_\rho$ is analogous to 
the quantity $\epsilon_P$ of Eq.~(\ref{disk}).  
The asymptotic behavior,  
\begin{equation}
{Disc~ {\cal M}^{(\rho)}_{ B^0 \to \rho^0 \gamma}\over 
{\cal M}_{ B^0 \to \rho^+ \rho^-}} 
\sim \left( {s\over s_0}\right)^{\alpha_\rho^0 -1} \ \ ,
\label{asym}
\end{equation}
justifies in this case use of an unsubtracted dispersion relation 
for the amplitude, 
\begin{eqnarray} \label{rho}
{\cal M}^{(\rho)}_{ B^0 \to \rho^0 \gamma}(m_B^2) &=& 
\frac{\epsilon_\rho}{\pi} {\cal M}_{B^0 \to \rho^+ \rho^-} 
{1\over \pi}  \int^\infty_{4 m_\rho^2} ds \ 
 \left({s\over s_0}\right)^{\alpha_\rho^0 - 1}~
{1\over s - m_B^2} \nonumber \\
&\simeq& \frac{\epsilon_\rho}{\pi} {\cal M}_{B \to \rho^+ \rho^-} 
\int^\infty_{4 m_\rho^2} \frac{ds}{\sqrt{s} (s - m_B^2)} 
= \frac{\epsilon_\rho}{\pi} { \sqrt{s_0} \over m_B} 
\ln \frac{2 m_\rho - m_B}
{2 m_\rho + m_B} ~
{\cal M}_{B^0 \to \rho^+ \rho^-} \ \ , 
\end{eqnarray}
where we approximate $\alpha_\rho^0 \simeq 0.5$.  
The above discussion is extendable in like manner 
to $B^+ \to \rho^+ \gamma$ decay.  

\section{The Weak Decay Vertex}

\noindent
The most general form for the weak amplitude $B \to \rho \rho$ 
is$^{\cite{gv}}$  
\begin{equation} \label{weakme}
{\cal M}_{B(p) \to \rho (k_1) \rho (k_2)} = 
\epsilon^{*\mu} (k_2) \epsilon^{*\nu} (k_1)
\left[ 
a g_{\mu \nu} + \frac{b}{m_\rho^2} p_\mu p_\nu +
i \frac{c}{m_\rho^2} \epsilon_{\mu \nu \alpha \beta} k_1^\alpha
p^\beta \right] \ \ .  
\end{equation}
The quantities $a,b,c$ can be interpreted as partial-wave 
amplitudes as they exhibit the respective threshold 
behavior of $S,D,P$-waves.  To proceed further requires knowledge 
of $\{ a, b, c\}$ for both the $B^0\to \rho^0 \rho^0$ and 
$B^0\to \rho^+ \rho^-$ weak decays.  Since no data yet exists for 
the $B \to \rho \rho$ transitions, we must determine $\{ a, b, c\}$ 
theoretically.   We have employed the BSW description of the 
nonleptonic $B$ transitions.$^{\cite{bsw}}$  In this model, 
the amplitude for the $B^0 \rho^+ \rho^-$ transition is 
\begin{equation}
{\cal M}_{B^0(p) \to \rho^+(k_1) \rho^-(k_2)} =
\frac{G_F}{\sqrt{2}} V_{ub} V_{ud}^* ~a_1 ~
\langle \rho^- (k_2) | \bar d \gamma^\mu u | 0 \rangle
\langle \rho^+ (k_1)| \bar u \gamma_\mu (1+\gamma_5) b | B^0 (p) \rangle \ \ ,
\end{equation}
where $a_1 \simeq 1.03$.$^{\cite{bhp}}$    
We adopt standard notation for the matrix elements, 
\begin{eqnarray}
\langle \rho^- (k_2)| \bar u \gamma^\mu d | 0 \rangle &=&
f_\rho m_\rho \epsilon^{*\mu} (k_2) \nonumber \\
\langle \rho^+ (k_1)| \bar u \gamma_\mu  b | B^0 (p)\rangle &=& 
\frac{2i}{m_B + m_\rho} \epsilon_{\mu \nu \alpha \beta}
\epsilon^{*\nu}(k_1) k_1^\alpha p^\beta ~ V (k_2^2) \label{ff0} \\
\langle \rho^+ (k_1)| \bar u \gamma_\mu \gamma_5 b | B^0 (p) \rangle &=&
(m_B + m_\rho) \epsilon^*_\mu (k_1) A_1 (k^2_2) - 
\frac{\epsilon^* (k_1) \cdot k_2}{m_B + m_\rho}
(p + k_1)_\mu A_2 (k_2^2) \nonumber \\
&-& \frac{2 \epsilon^* (k_1) \cdot k_2}{k_2^2} {k_2}_\mu m_\rho
\Bigr [ A_3 (k_2^2) - A_0 (k_2^2) \Bigr ]  \ \ .
\nonumber 
\end{eqnarray}
In what follows we assume the $k_2^2$ behavior of
the form factors to be of the simple pole form, and for 
$k_2^2 = m_\rho^2$ we obtain 
\begin{eqnarray}
V(k_2^2) &=& {h_V \over 1 - k_2^2 /m_{B^*}^2} \simeq h_V 
\nonumber \\
A_i (k_2^2) &=& {h_{A_i} \over 1 - k_2^2 /m_{B^*}^2} \simeq h_{A_i} 
\quad (i = 0,\dots,3) \ , 
\label{pole}
\end{eqnarray}
with $h_V \simeq 0.33, h_{A_1}\simeq h_{A_2} \simeq 
0.28 $.$^{\cite{bsw}}$ 
This gives for the amplitudes in Eq.~(\ref{weakme}), 
\begin{eqnarray} 
a &\simeq& \frac{G_F}{\sqrt{2}} V_{ub} V_{ud}^* m_\rho
a_1 f_\rho ( m_B + m_\rho) h_{A_1} \ ,\nonumber \\
b &\simeq& - \frac{G_F}{\sqrt{2}} V_{ub} V_{ud}^*
a_1 f_\rho \frac{2m_\rho^3}{m_B + m_\rho} h_{A_2} \ ,\label{ff} \\
c &\simeq& \frac{G_F}{\sqrt{2}} V_{ub} V_{ud}^*
a_1 f_\rho \frac{2m_\rho^3}{m_B + m_\rho} h_V \ \ . \nonumber 
\end{eqnarray}
The set of relevant helicity amplitudes can be written using explicit
form of the polarization vectors. The connection between the 
transverse helicity amplitudes and form factors in 
Eq.~(\ref{weakme}) is$^{\cite{gp}}$  
\begin{equation} \label{helamp}
{\cal M}^{++}_{\rho\rho} = a + \sqrt{x^2-1} ~c \ ,\qquad 
{\cal M}^{--}_{\rho\rho} = a - \sqrt{x^2-1} ~c \label{relation} 
\end{equation}
where $x \equiv (m_B^2 - 2 m_\rho^2)/2 m_\rho^2$. For 
completeness, we note that branching
ratios for the $B \to \rho \rho$ decays calculated from 
Eq.~(\ref{helamp}) amount to
${\cal B}_{\bar B^0 \to \rho^0 \rho^0} = 4.8 \cdot 10^{-7}$, 
${\cal B}_{\bar B^0 \to \rho^+ \rho^-} = 2.1 \cdot 10^{-5}$ and 
${\cal B}_{\bar B^- \to \rho^- \rho^0} = 1.5 \cdot 10^{-5}$, 
where all three $\rho\rho$ helicity configurations 
have been summed over.  

\section{Numerical Results}

Before numerically estimating the long-distance effects in $B\to \rho\gamma$, 
we first recall the determination of the short-distance contribution 
which arises from the effective Hamiltonian$^{\cite{lcsr}}$   
\begin{eqnarray}
{\cal H}^{\rm s.d.} &=& \frac{G_F}{\sqrt{2}} \Bigl( 
V_{tb} V_{td}^* C_7^{\mbox{\small{eff}}}~ O_7 + \dots  \Bigr) \ \ ,
\nonumber \\
O_7 &=& \frac{em_b}{8 \pi^2} \bar d \sigma_{\mu \nu}
(1 - \gamma_5) F^{\mu \nu} b \ \ ,
\end{eqnarray}
where $C_7^{\mbox{eff}} = -0.306$.  The amplitude 
for the short-distance $B \to \rho \gamma$ contribution is 
\begin{eqnarray}
\lefteqn{{\cal M}_{\rm s.d.}^{\lambda_\gamma, \lambda_\rho} =} \nonumber \\
& & - \frac{G_F}{\sqrt{2}} V_{tb} V_{td}^* C_7^{\mbox{\small{eff}}}~
\frac{em_b}{4 \pi^2}~ \epsilon^{*\mu} (q, \lambda_\gamma) 
\epsilon^{*\nu}(k_1, \lambda_\rho) ~2 F^{\rm S} (0) 
\left[
\epsilon_{\mu \nu \alpha \beta} k_1^\alpha p^\beta -
i \left( m_B E_\gamma g_{\mu \nu} - p_\mu q_\nu \right) \right] \ ,
\end{eqnarray}
where $\lambda_{\gamma}, \lambda_\rho$ are the respective 
photon, rho helicities and 
$F^{\rm S} (q^2 = 0)$ is a form factor related to the 
$B$-to-$\rho$ matrix element of $O_7$.  It is estimated from 
QCD sum rules$^{\cite{lcsr}}$ that 
$F^{\rm S} (0) \simeq 0.17$ for $B^0 \to \rho^0 \gamma$ 
and $F^{\rm S} (0) \simeq 0.17 \sqrt{2}$ for $B^- \to \rho^- \gamma$.   
This $B \to \rho\gamma$ matrix element can be written in a form
similar to Eq.~(\ref{weakme}) with the identification
\begin{eqnarray}
a^{\rm s.d.} &=& i \frac{G_F}{\sqrt{2}} 
V_{tb} V_{td}^* C_7^{\mbox{\small{eff}}}~
\frac{e m_b}{2 \pi^2} m_B E_\gamma F^S (0) \nonumber \\
c^{\rm s.d.} &=& -b^{\rm s.d.} = i \frac{G_F}{\sqrt{2}} 
m_\rho^2 V_{tb} V_{td}^* C_7^{\mbox{\small{eff}}}~
\frac{e m_b}{2 \pi^2} F^S (0) \ \ .
\end{eqnarray}
These can be related to the set of helicity amplitudes by
using Eq.~(\ref{relation}) which then 
contribute to the decay rate for $B \to \rho \gamma$ as 
\begin{equation}
{\Gamma_{B \to \rho \gamma}}^{\rm s.d.} = \frac{|{\bf q}|}{8 \pi m_B^2} 
\Bigl ( |{\cal M}^{++}_{s.d.}|^2 + 
|{\cal M}^{--}_{s.d.}|^2 \Bigr ) \ \ ,
\end{equation}
where ${\bf q}$ is the photon momentum.  

In our numerical work, we have adopted for the CKM matrix 
elements the values$^{\cite{al}}$ 
\begin{equation}
|V_{cb}| = 0.0393\ , \quad |V_{ub}| = 0.08~|V_{cb}| \ , \quad 
1.4 \le \bigg| {V_{td} \over V_{ub}} \bigg| \le 4.6 \ .
\label{ckm}
\end{equation}
The biggest source of uncertainty for the short-distance 
contribution is clearly the magnitude of $V_{td}$.  The 
above range leads to a range of branching ratios for the 
short distance component 
\begin{equation}
{\cal B}^{\rm s.d.}_{B^0 \to \rho^0 \gamma} \equiv 
{\Gamma^{\rm s.d.}_{B^0 \to \rho^0 \gamma}\over 
\Gamma^{\rm tot.}_{B^0}}  \simeq 10^{-7} \longrightarrow 10^{-6} \ \ , 
\label{brshort}
\end{equation}
with ${\cal B}^{\rm s.d.}_{B^- \to \rho^- \gamma} = 2
{\cal B}^{\rm s.d.}_{B^0 \to \rho^0 \gamma}$.  
It is this large theoretical spread which has motivated 
an experimental determination of ${\cal B}_{B \to \rho \gamma}$ 
as perhaps the best solution to the $V_{td}$ problem.  
At present, however, even the largest theoretical values are 
still considerably smaller than the experimental 
upper bounds ${\cal B}_{B^0 \to \rho^0 \gamma} \le 
3.9 \times 10^{-5}$ and ${\cal B}_{B^- \to \rho^- \gamma} \le 
1.1 \times 10^{-5}$.$^{\cite{patt}}$  

Upon including both short-distance and long-distance 
contributions, we obtain 
\begin{equation}
{\Gamma_{B \to \rho \gamma}} =
\frac{|{\bf q}|}{8 \pi m_B^2}
\Bigl ( |{\cal M}_{\rm s.d.}^{++} + 
{\cal M}_{\rm l.d.}^{++}|^2 + 
|{\cal M}_{\rm s.d.}^{--} + 
{\cal M}_{\rm l.d.}^{--}|^2 \Bigr ) \ \ .
\end{equation}
The long-distance amplitudes are in turn 
summed over contributions from the Pomeron and $\rho$ trajectories,  
\begin{equation}
{\cal M}_{\rm l.d.}^{\lambda_1 \lambda_2} =
\beta^{(P)} {\cal M}^{\lambda_1 \lambda_2}_{\rho^0 \rho^0}+
\beta^{(\rho)} {\cal M}^{\lambda_1 \lambda_2}_{\rho^+ \rho^-} \ \ .
\label{bar}
\end{equation}
Using Eqs.~(8),(11) and dropping the subtraction
constant but keeping the terms which are clearly related to
rescattering, we obtain
\begin{equation} \label{coeff}
\beta^{(P)} = \frac{\epsilon_P}{\pi} 
\ln \Bigl( 1 - \frac{m_B^2}{4 m_\rho^2} \Bigl) \ ,
\qquad 
\beta^{(\rho)} = \frac{\epsilon_\rho}{\pi}{\sqrt{s_0} \over m_B}
\ln \frac{2 m_\rho - m_B} {2 m_\rho + m_B} \ \ .
\end{equation}

Also needed for estimation of the FSI 
effect are the residue functions $\gamma_P$ and 
$\gamma_\rho$.   For the Pomeron case, we use experimental data on 
photoproduction of $\rho^0$ mesons from a nucleon 
target.$^{\cite{photo}}$  Upon calculating the differential 
cross section $d \sigma / dt$ at $t \simeq 0$ from the matrix element
of Eq.~(\ref{regge}) and applying the quark-counting rule 
to relate Pomeron-$pp$ and Pomeron-$\rho \rho$ couplings, 
we obtain $\gamma_P \simeq 4.53$.  Actually we use twice 
this value as the final-state photon can arise from either 
scattering vertex.  For $\rho$ exchange, 
we use $\gamma p \to \rho^+ n$ data$^{\cite{rhopl}}$ together 
with the isospin relation $\gamma_\rho (\gamma \rho^0 \to
\rho^+ \rho^-)/ \gamma_\rho(\gamma p \to \rho^+ n) \simeq 2$ 
to obtain $\gamma_\rho \simeq 7.6$.  

The recent review of the CKM matrix given in Ref.~\cite{al} cites 
$\rho = 0.05$ and $\eta = 0.36$ as providing the best fit to current 
data, where $\rho$ and $\eta$ are the Wolfenstein parameters
\begin{equation}
V_{ub} = A\lambda^3 (\rho - i \eta) \ , \qquad 
V_{td} = A\lambda^3 (1 - \rho - i \eta) \ \ , 
\label{wolf}
\end{equation}
and which corresponds to $| V_{td} / V_{ub}| \simeq 2.8$.
Assuming these central values, we obtain for the short distance and 
short-plus-long distance branching ratios the respective values 
\begin{equation}
{\cal B}^{\rm s.d.}_{B^0 \to \rho^0 \gamma} \simeq 7.3 \times 10^{-7} 
\ , \qquad {\cal B}^{\rm tot}_{B^0 \to \rho^0 \gamma} \simeq 
7.8 \times 10^{-7} \ \ .
\label{result}
\end{equation}
Figure~2 displays the effect of varying the 
$\rho$, $\eta$ values over their allowed range$^{\cite{al}}$.  
The upper and lower curves define the 
band of allowed values for the ratio 
$({\cal B}^{\rm tot}_{B^0 \to \rho^0 \gamma} -
{\cal B}^{\rm s.d.}_{B^0 \to \rho^0 \gamma})/ 
{\cal B}^{\rm s.d.}_{B^0 \to \rho^0 \gamma}$.  The upper 
curve is seen to be about $8\%$ for virtually the 
entire physical range of $\rho$.  

\section{Concluding Remarks}

\noindent
Final state rescattering effects may sometimes, although not always,
modify the usual analysis of decay processes. The situations where
rescattering can be important are those where there is a copiously 
produced final state which can rescatter to produce the decay mode being
studied. In our case the most important channel is $B \to
\rho^+\rho^-$, which is color allowed. It is also proportional to 
$V_{ub}$ instead of $V_{td}$ so that it is clearly a distinct 
contribution, and one that becomes more important if $V_{td}$ is at 
the lower end of its allowed range. Our results should also be 
adjusted upward or downward if the measured branching ratio proves 
to be larger or smaller than that predicted above using the BSW model.

The magnitude of the $\rho\rho \to \rho \gamma$ soft-scattering 
for $B^0 \to \rho^0 \gamma$, summarized 
in Eq.~(\ref{result}), is seen to occur at about 
the $8\%$ level.  This differs from 
the QCD sum rule estimates of Ref.~\cite{lcsr} which found 
a $10\%$ long distance contribution to ${\cal B}_{B^+ \to \rho^+ \gamma}$ 
but only a $1\%$ effect for ${\cal B}_{B^0 \to \rho^0 \gamma}$. 
Our result is especially noteworthy in view of the many other possible 
contributions to the unitarity sum and the finding of Ref.~\cite{we}
that multi-particle intermediate states are likely to dominate. 
Thus, although it is not possible at this time for anyone to 
completely analyze the long distance component, it seems plausible 
that the FSI contribution has the potential to occur 
at the $10\%$ level and perhaps even higher.\footnote{However, we 
do not claim the dominance of the long-distance contribution 
over the short-distance amplitude, as was done in 
Ref.~\cite{blok} for $B \to \pi\pi$.}  We concur 
with comments in the literature$^{\cite{gp},\cite{lcsr},\cite{aap}}$ 
that a deviation from isospin relations based on the short distance 
amplitude ({\it e.g.} ${\cal B}_{B^- \to \rho^- \gamma} = 
2{\cal B}_{B^0 \to \rho^0 \gamma}$) will be evidence for a 
long distance component, and such is the case here. 
Defining a parameter $\Delta$ to measure the isospin violation as
\begin{equation}
\Delta \equiv 1 - {1\over 2} {\Gamma_{B^- \to \rho^- \gamma} \over 
\Gamma_{B^0 \to \rho^0 \gamma}} \ \ ,
\end{equation}
we display in Fig.~3 the dependence of the effect upon the 
Wolfenstein parameter $\rho$. The isospin violation is found 
to be largest for large $\rho$.  

The FSI mechanism described here will not markedly affect the 
$B \to K^* \gamma$ rate.  The leading CKM contribution would involve 
a $D_s^* - {\bar D}^*$ intermediate state and thus be 
very suppressed since the soft-scattering requires 
a Reggeon carrying the quantum numbers of the $D$ meson.  
The non-leading CKM amplitudes, involving 
contributions from light-quark intermediate 
states, are proportional to $V_{ub}$ and are likewise very 
suppressed.  

Finally, we point out the implications of our work 
to obtaining a CP-violating signal.  Since $B \to \rho 
\rho$ decay is governed by 
CKM matrix elements which differ from those describing the short distance 
$B \to \rho \gamma$ transition, the necessary condition for CP violation 
is satisfied. We consider the CP asymmetry 
\begin{equation}
a = \frac{ | \Gamma_{B \to \rho \gamma} -
\Gamma_{\bar B \to \rho \gamma} | }
{\Gamma_{B \to \rho \gamma} +
\Gamma_{\bar B \to \rho \gamma}} =
\frac{|{\cal B}_{B \to \rho \gamma}-
{\cal B}_{\bar B \to \rho \gamma}|}
{{\cal B}_{B \to \rho \gamma}+
{\cal B}_{\bar B \to \rho \gamma}} \ \ .
\end{equation}
We have studied the resulting effect numerically. 
The FSI are found to increase the amplitude for 
$B^0 \to \rho^0 \gamma$ but reduce it for ${\overline B}^0 
\to \rho^0 \gamma$.  This is what gives rise to the 
asymmetry and we find $a \simeq 7 \%$.  As with our 
other results, we take this magnitude as an indication 
that experimentally interesting signals might well exist 
and should not be ignored in planning for future studies.

\noindent

We would like to thank Jo\~ao M. Soares for reading the manuscript and 
useful comments. This work was supported in part by the U.S. 
National Science Foundation.

%
%
%
%
%

\begin{figure}[t]
\centering
\centerline{
\epsfbox{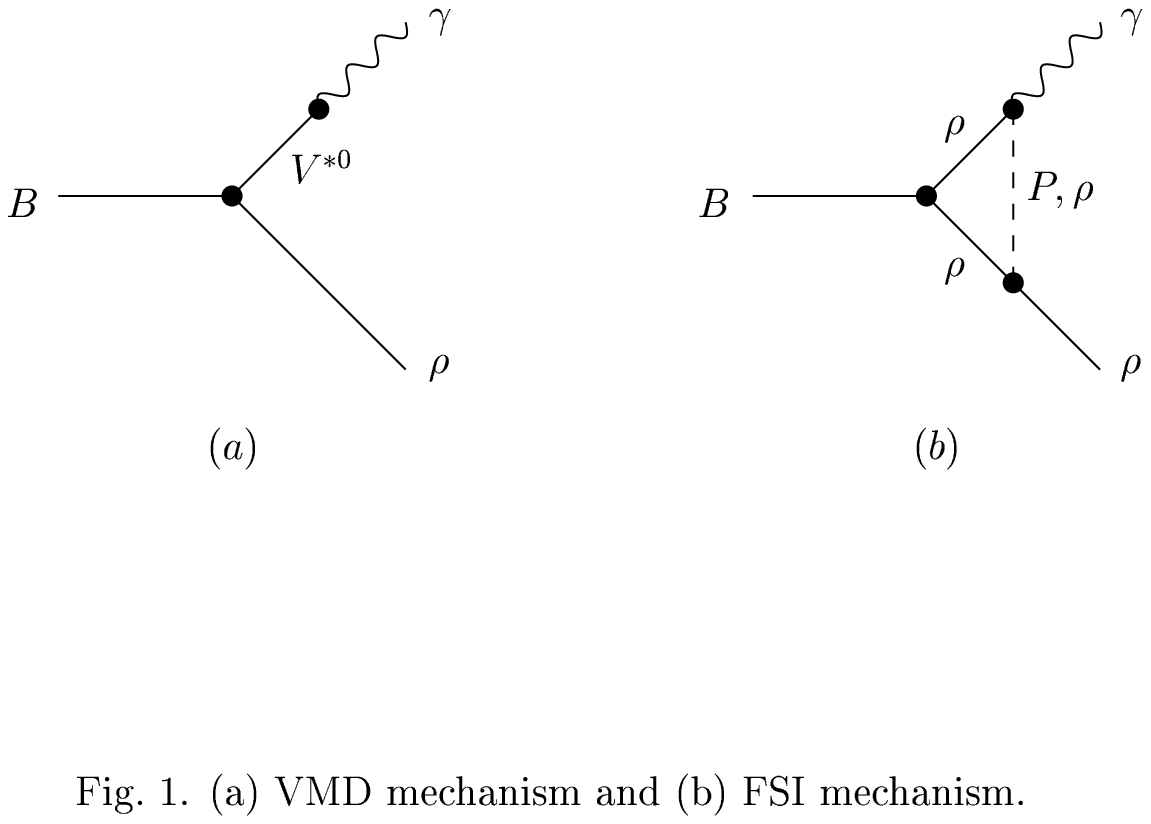}}
\caption{$B\to \rho\gamma$ via (a) Vector dominance
process and (b) final state interaction effect.}
\end{figure}

\begin{figure}[t]
\centering
\leavevmode
\centerline{
\epsfbox{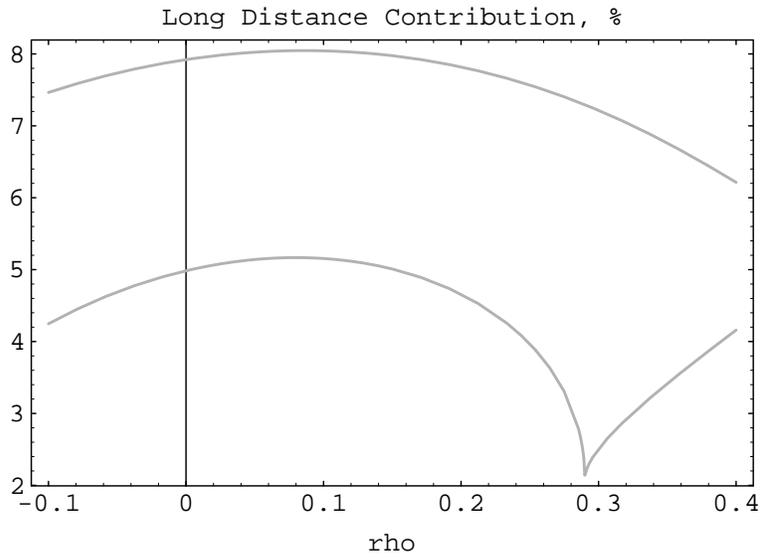}}
\caption{Dependence of the FSI effect upon the CKM
parameter $\rho$.}
\end{figure}

\begin{figure}[t]
\centering
\leavevmode
\centerline{
\epsfbox{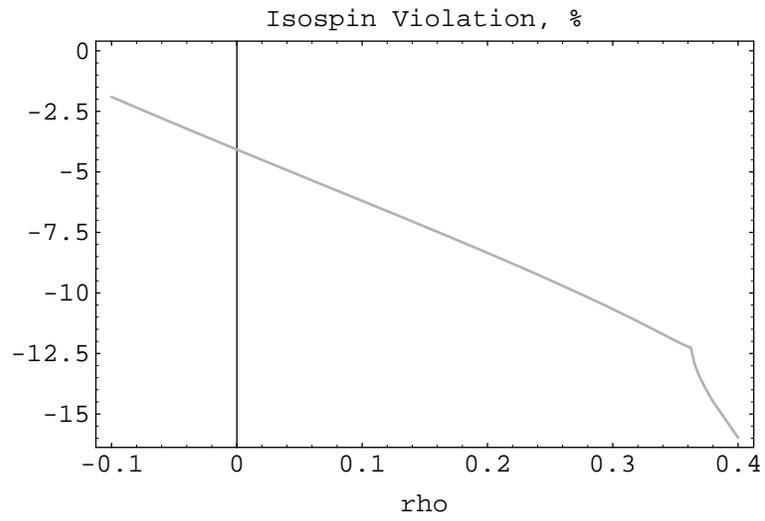}}
\caption{Dependence of isospin violation upon the CKM
parameter $\rho$.}
\end{figure}

\end{document}